\documentclass[3p,10pt]{elsarticle}

\usepackage{amsmath}
\usepackage{amssymb}
\usepackage{amsthm}
\usepackage{algorithm}
\usepackage[noend]{algpseudocode}
\usepackage[hidelinks]{hyperref}

\newtheorem{lemma}{Lemma}
\newtheorem{theorem}{Theorem}

\newtheorem{definition}{Definition}
\newtheorem{problem}{Problem}

\newcommand{\tisr}{\textsc{Temporal Independent Set Reconfiguration}}

\journal{Information Processing Letters}

\begin{document}

\begin{frontmatter}

\title{$k$'s a crowd: Reconfiguring independent sets in temporal graphs}

\author[st-andrews]{Duncan Adamson\corref{cor1}}
\ead{duncan.adamson@st-andrews.ac.uk}
\cortext[cor1]{Corresponding author}

\affiliation[st-andrews]{organization={Department of Computer Science, University of St Andrews},
                         addressline={North Haugh},
                         postcode={KY18 9SX},
                         country={United Kingdom}}

\begin{abstract}
In \tisr{}, we seek the maximum number of tokens that can occupy distinct, pairwise non-adjacent vertices in every snapshot of a temporal graph. To obtain the next set, each token may stay in place or slide along one edge of the current snapshot.

In this paper, we show that we can find, in $O(nT)$ time, a maximum reconfigurable sequence on temporal path graphs of $n$ vertices. Further, we provide a new hardness bound, showing that deciding whether a reconfigurable sequence of size $k$ exists is NP-complete even when the underlying graph is a tree of depth two.
\end{abstract}

\end{frontmatter}

\section{Introduction}

\emph{Independent sets} are widely studied in computer science and graph theory~\cite{10.1007/978-3-032-26465-7_2,Balliu2021,Enright2027,Hermelin2023,Meeks2022,Mertzios2023}. Independent sets and related problems, such as graph colouring and matching, have applications in epidemiology, resource allocation, and cybersecurity. In these settings, maintaining independence can prevent two agents from competing for the same resource at the same time or coming into contact in a way that allows a virus to spread.

Davot, Enright and Larios-Jones~\cite{Davot2026Reconfiguration} introduced a framework for reconfiguring vertex sets in \emph{temporal graphs}. A temporal graph is a sequence of static graphs $G_1,\ldots,G_T$, called \emph{snapshots}, over a common vertex set. In the \tisr{} problem, we seek a sequence $(S_1,\ldots,S_T)$ of independent sets, one in each snapshot, that accommodates as many tokens as possible. For each $t<T$, a token on a vertex of $S_t$ can stay in place or slide along an edge of $G_t$. The resulting set $S_{t+1}$ must be independent in $G_{t+1}$. Following a single token through these reconfigurations gives a temporal walk with at most one move per snapshot, as in temporal graph exploration~\cite{adamson2026exploration,erlebach2021temporal}.

Davot et al.~\cite{Davot2026Reconfiguration} give a fixed-parameter algorithm for the combined parameter of lifetime, defined as the number of snapshots in the temporal graph, and treewidth of the \emph{underlying graph}, the graph formed by the union of the snapshots. We consider the case of unrestricted lifetime, noting that even for an underlying graph of treewidth one, the running time of their algorithm depends on the lifetime through its parameter function.

\paragraph*{Our results}
On a temporal path with $n$ vertices and lifetime $T$, we construct a maximum reconfigurable sequence and its bijections in $O(nT)$ time by choosing the leftmost feasible walk for each successive token. When the underlying graph is a tree of depth two, deciding whether a sequence of size $k$ exists is NP-complete.

\paragraph*{Related work}
Several notions of disjoint temporal walks restrict how the walks can use the same graph. Ibiapina and Silva~\cite{Ibiapina2024SnapshotDisjointness} study snapshot disjointness, which requires different walks to traverse edges at disjoint sets of times. In a reconfigurable sequence, several tokens may slide along edges of $G_t$ at the same time. Their destinations must be distinct and pairwise non-adjacent in $G_{t+1}$.

In independent set reconfiguration under token sliding, each step moves one token along an edge of a static graph and preserves independence. Demaine et al.~\cite{Demaine2015SlidingTokens} give a linear-time algorithm for deciding whether such a sequence connects two given independent sets on a tree. Hoang~\cite{Hoang2024DistanceIndependentSet} studies the corresponding reconfiguration problem when tokens must maintain a prescribed minimum pairwise distance. Hirahara et al.~\cite{Hirahara2025Extended} allow several tokens to slide simultaneously on a fixed graph. They ask whether one prescribed independent set can reach another when at most a given number of tokens move in each step. In the temporal problem, the available slides and the independent sets both change with the snapshot.

The movement rule also resembles classical multi-agent pathfinding, where agents can wait or move to neighbours as they travel between prescribed start and goal vertices, subject to conflict constraints~\cite{Stern2019MAPF}. In \tisr{}, the initial and final independent sets are free. The objective is to maximise the number of tokens that can be retained throughout the sequence.

Multistage graph problems also couple solutions at consecutive snapshots. Fluschnik et al.~\cite{Fluschnik2022MultistageVertexCover} bound the symmetric difference between successive vertex covers. Chimani et al.~\cite{Chimani2023Multistage} study how much overlap can be retained between successive snapshot-optimal solutions, including for independent set on bipartite or planar graphs. In our problem, a transition is feasible when a bijection pairs each currently occupied vertex with a distinct next position reachable by waiting or by traversing an active edge. Davot et al.~\cite{Davot2026Reconfiguration} test for such a reconfiguration bijection by finding a perfect matching between copies of the two sets.

Hermelin et al.~\cite{Hermelin2023} studied temporal $\Delta$-independent sets. They seek a fixed vertex set in which each pair is non-adjacent in at least one snapshot of every window of $\Delta$ consecutive snapshots. They showed that the decision problem is NP-complete even on \emph{temporal interval graphs}, in which each snapshot is an interval graph. On a static interval graph, a maximum independent set can be found in polynomial time. In our model, the occupied vertices can change as the tokens move.

Mertzios et al.~\cite{Mertzios2023} studied \emph{maximum temporal matching}. They seek a largest set of time-stamped edges such that each vertex is matched at most once in any window of $\Delta$ consecutive snapshots. In the line graph, choosing an edge corresponds to choosing a vertex at a time when the edge is active. Selections in the same window must correspond to edges with disjoint endpoints. The problem is NP-hard even when the underlying graph is a path. In our setting, each token's temporal walk also constrains which vertex it can occupy next.

\paragraph*{AI disclosure statement.}
ChatGPT Astra was used in polishing and proofreading this work. All results were human-derived, and the author maintains full accountability for the presented work.

\section{Preliminaries}
\label{sec:prelims}

For $i,j\in\mathbb{N}$, let $[i,j]=\{i,i+1,\ldots,j\}$, with $[i,j]=\emptyset$ when $i>j$. We write $[i]$ for $[1,i]$.

Let $G=(V,E)$ be a simple undirected graph with vertex set $V=\{v_1,\ldots,v_n\}$ and edge set $E=\{e_1,\ldots,e_m\}$. We write $(u,v)$ for the edge with endpoints $u$ and $v$, in either order. A \emph{walk} is a sequence of vertices in which each consecutive pair is joined by an edge. Its length is the number of edges traversed, counting repetitions. A graph is \emph{connected} if a walk joins every pair of vertices.

The \emph{neighbourhood} of $v$ is $N(v)=\{u\in V\mid (u,v)\in E\}$. The \emph{degree} of $v$ is $\Delta(v)=|N(v)|$. Two vertices $u$ and $v$ are adjacent if $(u,v)\in E$. A set $I\subseteq V$ is \emph{independent} in $G$ if its vertices are pairwise non-adjacent.

A \emph{temporal graph} $\mathcal{G}=(G_1,\ldots,G_T)$ of \emph{lifetime} $T\geq 1$ consists of $T$ ordered \emph{snapshots} $G_t=(V,E_t)$ over a common vertex set $V$. An edge is \emph{active} in snapshot $t$ if it belongs to $E_t$. We write $N_t(v)$ for the neighbourhood of $v$ in $G_t$. The \emph{underlying graph} of the temporal graph $\mathcal{G}$ is the graph $U(\mathcal{G})=(V,\bigcup_{t=1}^T E_t)$. When referring to $\mathcal{G}$, we use $\Delta(v)$ for the degree of $v$ in its underlying graph, and $\Delta_t(v)$ for the degree of $v$ in $G_t$.

\paragraph*{Reconfigurable sequences}
Following Davot et al.~\cite{Davot2026Reconfiguration}, we say that $S\subseteq V$ is \emph{reconfigurable into} $S'\subseteq V$ at time-step $t$ if there is a bijection $f:S\to S'$ such that, for each $v\in S$, either $f(v)=v$ or $(v,f(v))\in E_t$. We call $f$ a \emph{reconfiguration bijection}. A token placed at $v$ stays in place in the first case and slides to $f(v)$ in the second.

A \emph{reconfigurable sequence of independent sets} is a sequence $(S_1,\ldots,S_T)$ such that $S_t$ is independent in $G_t$ for each $t\in[T]$, and $S_t$ is reconfigurable into $S_{t+1}$ at each $t<T$. Its \emph{size} is $|S_1|$, since all sets in the sequence have the same cardinality. Where independence is understood, we simply write \emph{reconfigurable sequence}.

\begin{figure}
    \centering
    \includegraphics[width=0.5\linewidth]{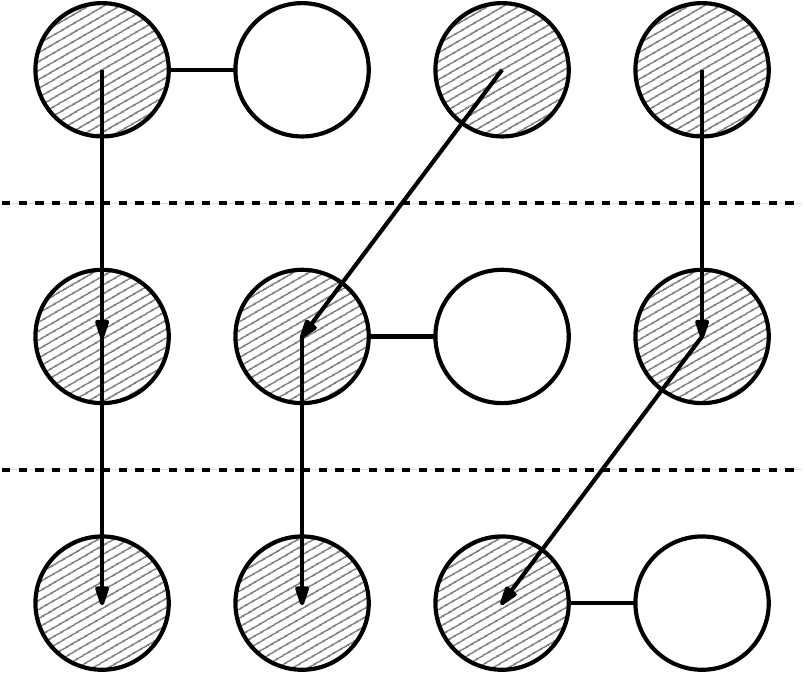}
    \caption{A reconfigurable sequence of size three on a temporal path. Each row shows a snapshot and its independent set (shaded). Arrows indicate the reconfiguration bijections between successive rows. A diagonal arrow represents a slide along an edge in the row it leaves. The underlying graph has independence number two.}
    \label{fig:temp_ind_walk_example}
\end{figure}

To describe individual tokens, label the $k$ tokens in $S_1$ from $1$ to $k$ and fix a reconfiguration bijection $f_t:S_t\to S_{t+1}$ for each $t\in[T-1]$. The temporal walk of token $i$ consists of its positions $W_i[1],\ldots,W_i[T]$, with $W_i[t+1]=f_t(W_i[t])$. Thus $S_t=\{W_i[t]\mid i\in[k]\}$ for a sequence of size $k$. Figure~\ref{fig:temp_ind_walk_example} gives an example.

\begin{problem}[\tisr{}, decision version]
    Given a temporal graph $\mathcal{G}$ and a nonnegative integer $k$, does $\mathcal{G}$ admit a reconfigurable sequence of independent sets of size $k$?
\end{problem}

The optimisation problem asks for a maximum reconfigurable sequence. A sequence of size at least $k$ gives one of size exactly $k$ by retaining any $k$ tokens throughout. The decision problem is NP-hard even for a single snapshot, when it is equivalent to \textsc{Independent Set}. It belongs to NP since we can verify independence and the reconfiguration bijections in polynomial time.

\section{Paths}

We first consider \tisr{} on \emph{temporal paths}, that is, temporal graphs whose underlying graph is a path. Write the vertices as $v_1,\ldots,v_n$ in path order, so the underlying graph edges are $(v_i,v_{i+1})$ for $i\in[n-1]$. For simplicity, we write $v_i<v_j$ if $i<j$ and say that $v_i$ is to the left of $v_j$.

Our approach assigns temporal walks to the tokens from left to right. We first show that the leftmost token can wait at $v_1$ throughout.

\begin{lemma}
    \label{lem:leftWall}
    If a temporal path admits a reconfigurable sequence of size $k\geq 1$, it admits one in which a token stays at $v_1$ throughout.
\end{lemma}

\begin{proof}
    Fix a reconfigurable sequence and its bijections, and label the tokens in left-to-right order in the first snapshot. Their order is preserved thereafter. To exchange order in a single snapshot, two tokens would have to traverse the same edge in opposite directions. Both endpoints would already be occupied in the snapshot in which the tokens slide, contradicting independence.

    Replace $W_1$ by the constant walk at $v_1$. If token $1$ already occupies $v_1$ in snapshot $t$, this leaves $S_t$ unchanged. Otherwise every other token lies strictly to the right of $W_1[t]\geq v_2$, so none occupies $v_1$ or its only possible neighbour, $v_2$. The resulting walks still realise a reconfigurable sequence.
\end{proof}

Once a temporal walk has been assigned to one token, its positions restrict the vertices available to the next token on its right. We express this restriction by a sequence of bounds, one for each snapshot.

\begin{definition}[$B$-bounded leftmost temporal walk]
    Let $B=(b_1,\ldots,b_T)$ be a sequence of vertices of a temporal path. A temporal walk $W$ is \emph{$B$-bounded} if $W[t]>b_t$ for every $t\in[T]$. A $B$-bounded walk $W'$ is \emph{leftmost} if, for every $B$-bounded temporal walk $W''$ and every $t\in[T]$, we have $W'[t]\leq W''[t]$.
\end{definition}

\begin{lemma}
    \label{lem:LBW_computation}
    After $O(nT)$ preprocessing of a temporal path, we can compute a $B$-bounded leftmost temporal walk, or determine that none exists, in $O(T)$ time for any given bound sequence $B$.
\end{lemma}

\begin{proof}
    For each snapshot $t$ and path edge $(v_i,v_{i+1})$, record whether the edge is active. This takes $O(nT)$ time and allows constant-time edge queries. Let $L_t(i)$ be the smallest index a token at $v_i$ can reach during snapshot $t$. It is $i-1$ if $i>1$ and $(v_{i-1},v_i)\in E_t$, and $i$ otherwise. The reachable indices form an interval containing $i$, since the token may wait.

    Write $b_t=v_{\beta_t}$. If any $\beta_t=n$, the bound cannot be met. Otherwise the constant walk at $v_n$ meets every bound. For each $t$, consider the positions a token can occupy in snapshot $t$ while meeting the bounds from $t$ through $T$. We claim that their indices form a suffix $[\ell_t,n]$. At $T$ this holds with $\ell_T=\beta_T+1$. Suppose it holds at $t+1$. A token at $v_i$ with $i\geq\ell_{t+1}$ can wait. If $\ell_{t+1}>1$, the only vertex further left from which the token can reach this suffix in one move is $v_{\ell_{t+1}-1}$, and this requires $(v_{\ell_{t+1}-1},v_{\ell_{t+1}})\in E_t$. Intersecting these possible positions with the bound at $t$ gives
    \[
        \ell_t=\max\{\beta_t+1,L_t(\ell_{t+1})\}.
    \]

    Place the token at $v_{\ell_1}$ in snapshot~$1$. For $2\leq t\leq T$, if it occupies $v_i$ in snapshot $t-1$ with $i\geq\ell_{t-1}$, the interval of indices it can reach using $G_{t-1}$ intersects $[\ell_t,n]$. The leftmost index in that intersection is $\max\{\ell_t,L_{t-1}(i)\}$. This choice is pointwise leftmost. Indeed, for each snapshot $s$, the function $L_s$ is nondecreasing in its vertex index. If another bounded walk occupies $v_j$ with $j\geq i$ in snapshot $t-1$, its next index is at least $\max\{\ell_t,L_{t-1}(j)\}\geq\max\{\ell_t,L_{t-1}(i)\}$. Since $\ell_1$ is the least feasible first index, induction gives the claim at every snapshot. See Algorithm~\ref{alg:left-bounded-path} for pseudocode.
\end{proof}

\begin{algorithm}
    \caption{Computing a $B$-bounded leftmost temporal walk}\label{alg:left-bounded-path}
    \begin{algorithmic}
        \Procedure{LeftmostWalk}{$\mathcal{G}, B=(v_{\beta_1},\ldots,v_{\beta_T})$}
            \If{$\beta_t=n$ for some $t\in[T]$}
                \State \Return $\emptyset$
            \EndIf
            \State $\ell_T\gets\beta_T+1$
            \For{$t=T-1$ \textbf{down to} $1$}
                \State $\ell_t\gets\max\{\beta_t+1,L_t(\ell_{t+1})\}$
            \EndFor
            \State $i\gets\ell_1$, \quad $W'[1]\gets v_i$
            \For{$t=2$ \textbf{to} $T$}
                \State $i\gets\max\{\ell_t,L_{t-1}(i)\}$
                \State $W'[t]\gets v_i$
            \EndFor
            \State \Return $W'$
        \EndProcedure
    \end{algorithmic}
\end{algorithm}

For a token following a temporal walk $W$, define its right boundary $B(W)=(b_1(W),\ldots,b_T(W))$. Write $W[t]=v_i$. Set $b_t(W)=v_{i+1}$ if $i<n$ and $(v_i,v_{i+1})\in E_t$, and set $b_t(W)=v_i$ otherwise. A token to its right must occupy a vertex strictly to the right of $b_t(W)$ in snapshot $t$.

\begin{lemma}
    \label{lem:LBW_use}
    Let $W_1,\ldots,W_k$ be the temporal walks of the tokens in a reconfigurable sequence on a temporal path, labelled from left to right. For each $i\in[2,k]$, replacing $W_i$ by a $B(W_{i-1})$-bounded leftmost temporal walk gives a reconfigurable sequence.
\end{lemma}

\begin{proof}
    Write $W_i'$ for the replacement walk. Since $W_i$ is $B(W_{i-1})$-bounded, $W_i'[t]\leq W_i[t]$ for every $t\in[T]$. Moving token $i$ weakly to the left in each snapshot preserves its separation from every token to its right. On the left, $W_i'[t]>b_t(W_{i-1})$ ensures that tokens $i$ and $i-1$ occupy distinct, non-adjacent vertices in $G_t$. Every other token to the left of token $i$ is at least two path edges away, since token $i-1$ lies between them. Thus the modified walks realise a reconfigurable sequence.
\end{proof}

\begin{theorem}
    On a temporal path with $n$ vertices and lifetime $T$, we can compute a maximum reconfigurable sequence of independent sets and its reconfiguration bijections in $O(nT)$ time.
\end{theorem}

\begin{proof}
    Assign the first token the constant walk at $v_1$. After assigning $W_i$, compute $B(W_i)$ and assign the next token a $B(W_i)$-bounded leftmost walk. Stop when no such walk exists. At each $t\in[T]$, the new token lies strictly to the right of $b_t(W_i)$, so its vertex is distinct from and non-adjacent to $W_i[t]$ in $G_t$. The other tokens lie further left, so the walks realise a reconfigurable sequence.

    To prove optimality, fix a maximum reconfigurable sequence of size $q$ together with its bijections. By Lemma~\ref{lem:leftWall}, we may take its first walk to be our constant walk. Suppose its first $i<q$ walks agree with the construction. Its $(i+1)$st walk witnesses that a $B(W_i)$-bounded walk exists. Lemma~\ref{lem:LBW_use} allows us to replace that walk by our leftmost choice while preserving feasibility. Thus the construction can assign at least $q$ walks, and feasibility ensures that it assigns at most $q$.

    The preprocessing in Lemma~\ref{lem:LBW_computation} is performed once. Computing a boundary and finding the next walk each take $O(T)$ time. There are at most $n$ tokens and one unsuccessful call, giving $O(nT)$ time in total.
\end{proof}

\section{Trees}

We reduce from \textsc{Independent Set} by encoding the choice of an input vertex in the leaves of a depth-two tree.

\begin{theorem}
    The decision version of \tisr{} is NP-complete even when the underlying graph is a tree of depth two.
\end{theorem}

\begin{proof}
    \emph{Construction.} Let $(G,k)$ be an instance of \textsc{Independent Set}, where $G=(V,E)$, $n=|V|$ and $1\leq k\leq n$. We construct a temporal graph whose underlying tree has root $c$. Each input vertex gives one branch of the tree, and each input edge gives a sequence of tests.

    For each $v\in V$, give $c$ a child named $v$ and attach $L=n+2$ leaves $\ell_{v,1},\ldots,\ell_{v,L}$ to $v$. Every edge $(v,\ell_{v,h})$, $h\in[L]$, is active in every snapshot. Let $R=nL+2$. For each input edge $e=(u,v)\in E$ and each $j\in[R]$, introduce three new vertices $x_{(e,j)},y_{(e,j)},z_{(e,j)}$, private to test $(e,j)$. In the underlying tree, $x_{(e,j)}$ and $z_{(e,j)}$ are children of $c$, and $y_{(e,j)}$ is a child of $x_{(e,j)}$. The vertices $u$ and $v$ are the same children of $c$ in every test involving those endpoints. Figure~\ref{fig:tree-construction} shows the construction for a single input edge.

    \begin{figure}[htbp]
        \centering
        \includegraphics[width=.84\linewidth]{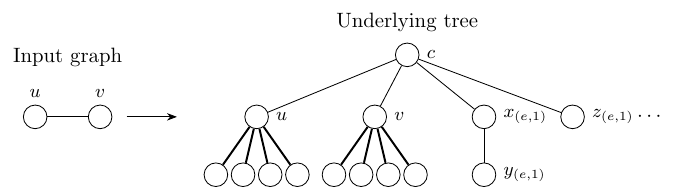}
        \caption{The construction for the input graph consisting of the single edge $e=(u,v)$. Each endpoint receives $L=4$ leaves. Thick edges are active in every snapshot. The vertices $x_{(e,1)},y_{(e,1)},z_{(e,1)}$ belong to the first test for $e$. The dots represent the other nine private triples attached to $c$. The input edge is tested through $c$.}
        \label{fig:tree-construction}
    \end{figure}

    Order the pairs $(e,j)\in E\times[R]$ arbitrarily. For each pair in this order, append the four snapshots $G_{(e,j),1},\ldots,G_{(e,j),4}$. For $e=(u,v)$, the table gives all active edges in addition to the permanent leaf edges.
    \[
    \begin{array}{c|l}
        \text{Snapshot} & \text{Additional active edges}\\ \hline
        G_{(e,j),1} & (c,x_{(e,j)})\\
        G_{(e,j),2} & (c,u),\ (c,v)\\
        G_{(e,j),3} & (c,u),\ (c,v),\ (c,z_{(e,j)})\\
        G_{(e,j),4} & (x_{(e,j)},y_{(e,j)}),\ (c,x_{(e,j)}).
    \end{array}
    \]
    After all tests, append a final snapshot $G_T$ containing exactly the permanent leaf edges and the edges $(c,v)$ for every $v\in V$. Let $P = 3R \vert E \vert$ be the number of private vertices, and $K = P + kL$ the target sequence size. The construction has $1+n+nL+P$ vertices and lifetime $T=4R|E|+1$, both polynomial in the size of $G$.

    \emph{Correctness.} 
    Fix $v\in V$ and distinct $a,b\in[L]$. Suppose $\ell_{v,a}$ and $\ell_{v,b}$ are occupied in snapshot $G_t$, where $t>1$. A token reaching either leaf from another vertex must have come from $v$ in $G_{t-1}$. If one token came from $v$, the other must have waited on its own leaf. This would place two tokens on adjacent vertices in $G_{t-1}$. Thus both tokens waited. By induction, two leaves occupied in $G_T$ remain occupied throughout, and their common neighbour $v$ is never occupied.

    Let $I\subseteq V$ be an independent set of size $k$ in $G$. In the first snapshot, place a token on every leaf $\ell_{v,h}$ with $v\in I$ and $h\in[L]$, and one on each private vertex. For each test $(e,j)$, where $e=(u,v)$ and $j\in[R]$, choose an endpoint $w\in\{u,v\}\setminus I$. Such an endpoint exists because $I$ is independent. The token at $x_{(e,j)}$ takes the following walk:
    \[
        x_{(e,j)},\ c,\ w,\ c,\ x_{(e,j)}.
    \]
    The first four positions are those in $G_{(e,j),1},\ldots,G_{(e,j),4}$, respectively. The fifth is the position in the snapshot immediately after $G_{(e,j),4}$. The table gives the departure edge for each slide. In $G_{(e,j),2}$ the token occupies $c$, whose only active neighbours are the empty vertices $u$ and $v$. In $G_{(e,j),3}$ it occupies $w$. Its neighbours there are $c$ and the leaves $\ell_{w,h}$, all empty since the token has left $c$ and $w\notin I$. In $G_{(e,j),4}$ the token again occupies $c$, whose only active neighbour is the empty vertex $x_{(e,j)}$. It then returns to $x_{(e,j)}$. All other tokens wait through these four moves. Each test therefore restores the initial placement, which is also independent in $G_T$. This gives a reconfigurable sequence of size $K$. Figure~\ref{fig:edge-test} illustrates one test.

    \begin{figure}[htbp]
        \centering
        \includegraphics[width=0.56\linewidth]{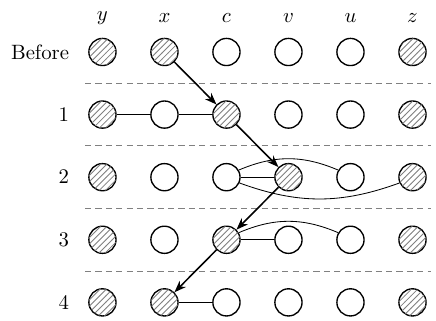}
        \caption{Test $(e,j)$ for $e=(u,v)$ with $u\in I$ and $v\notin I$. Row $r\in[4]$ shows $G_{(e,j),r}$ restricted to the six labelled vertices. Solid lines are active edges and hatching marks occupied vertices. The arrows trace the token initially at $x_{(e,j)}$. It uses $v$ in row~3 because the leaves of $v$ are empty. The last row shows only the positions in the following snapshot.}
        \label{fig:edge-test}
    \end{figure}

    Conversely, suppose there is a reconfigurable sequence of size $K$. Let $I$ consist of the vertices $v\in V$ with at least two occupied leaves in $G_T$. The preceding observation implies that no vertex of $I$ is occupied in any snapshot.

    In the final snapshot, each branch centred at a vertex of $I$ contains at most $L$ tokens. Every other branch centred at a vertex of $V$ contains at most one token, since at most one of its leaves is occupied and its centre cannot be occupied together with that leaf. If $|I|\leq k-1$, the total number of tokens, including those at the private vertices and the root, is at most
    \[
        P+1+|I|L+(n-|I|)\ \leq\ P+1+(k-1)L+n\ <\ P+kL.
    \]
    This contradicts the choice of $K$, so $|I|\geq k$.

    In $G_T$, the $n$ input branches and the root together contain at most $nL+1$ tokens. Of the $P$ private vertices, at most $nL+1-kL<R$ are therefore empty. Fix $e=(u,v)\in E$. Its $R$ private triples are disjoint, so there exists $j\in[R]$ for which $x_{(e,j)},y_{(e,j)},z_{(e,j)}$ are all occupied in $G_T$. Each of these vertices is isolated in every snapshot after $G_{(e,j),4}$. All three must therefore be occupied in the snapshot immediately following $G_{(e,j),4}$.

    Consider the reconfiguration bijections backwards from those three positions. Under the bijection from $G_{(e,j),4}$ to the following snapshot, the preimages of $x_{(e,j)}$ and $y_{(e,j)}$ must be two vertices of the active path $y_{(e,j)},x_{(e,j)},c$. Its only independent pair is $\{y_{(e,j)},c\}$. The predecessor of the token at $z_{(e,j)}$ is $z_{(e,j)}$ itself, so both $c$ and $z_{(e,j)}$ are occupied in $G_{(e,j),4}$. If $u,v\in I$, neither can be occupied in $G_{(e,j),3}$. The only possible predecessors of the tokens at $c$ and $z_{(e,j)}$ are then $c$ and $z_{(e,j)}$, which are adjacent in $G_{(e,j),3}$. This contradicts independence. Thus every edge of $G$ has an endpoint outside $I$, and $I$ is an independent set of size at least $k$. Any $k$ vertices of $I$ give a solution to the input instance.
\end{proof}

\bibliographystyle{elsarticle-num}
\bibliography{bib}

@Article{Hermelin2023,
  doi       = {10.1016/j.tcs.2023.113885},
  author    = {Hermelin, Danny and Itzhaki, Yuval and Molter, Hendrik and Niedermeier, Rolf},
  journal   = {Theoretical Computer Science},
  title     = {Temporal interval cliques and independent sets},
  year      = {2023},
  pages     = {113885},
  volume    = {961},
  publisher = {Elsevier},
}

@Article{Mertzios2023,
  doi       = {10.1016/j.jcss.2023.04.005},
  author    = {Mertzios, George B. and Molter, Hendrik and Niedermeier, Rolf and Zamaraev, Viktor and Zschoche, Philipp},
  journal   = {Journal of Computer and System Sciences},
  title     = {Computing maximum matchings in temporal graphs},
  year      = {2023},
  pages     = {1--19},
  volume    = {137},
  publisher = {Elsevier},
}

@Article{Enright2027,
  author    = {Enright, Jessica and Hand, Samuel D. and Larios-Jones, Laura and Meeks, Kitty},
  journal   = {Journal of Computer and System Sciences},
  title     = {Structural parameters for dense temporal graphs},
  year      = {2027},
  volume    = {163},
  doi       = {10.1016/j.jcss.2026.103846},
  pages     = {103846},
  publisher = {Elsevier},
}

@InProceedings{Balliu2021,
  doi       = {10.1145/3465084.3467901},
  author    = {Balliu, Alkida and Brandt, Sebastian and Kuhn, Fabian and Olivetti, Dennis},
  booktitle = {Proceedings of the 2021 ACM Symposium on Principles of Distributed Computing},
  title     = {Improved distributed lower bounds for {MIS} and bounded (out-)degree dominating sets in trees},
  year      = {2021},
  pages     = {283--293},
}

@InProceedings{Meeks2022,
  doi       = {10.1007/978-3-031-08740-0_16},
  author       = {Meeks, Kitty},
  booktitle    = {Revolutions and Revelations in Computability},
  title        = {Reducing reachability in temporal graphs: Towards a more realistic model of real-world spreading processes},
  year         = {2022},
  series       = {Lecture Notes in Computer Science},
  volume       = {13359},
  publisher    = {Springer},
  pages        = {186--195},
}

@InProceedings{10.1007/978-3-032-26465-7_2,
  doi       = {10.1007/978-3-032-26465-7_2},
author="Adamson, Duncan
and Mertzios, George B.
and Spirakis, Paul G.",
editor="Georgiou, Chryssis",
title="Maintaining Bipartite Colourings on Temporal Graphs on a Budget",
booktitle="Structural Information and Communication Complexity",
year="2026",
publisher="Springer Nature Switzerland",
address="Cham",
pages="21--33",
isbn="978-3-032-26465-7"
}

@Article{erlebach2021temporal,
  doi       = {10.1016/j.jcss.2021.01.005},
  author    = {Erlebach, Thomas and Hoffmann, Michael and Kammer, Frank},
  journal   = {Journal of Computer and System Sciences},
  title     = {On temporal graph exploration},
  year      = {2021},
  pages     = {1--18},
  volume    = {119},
  publisher = {Elsevier},
}

@inproceedings{adamson2026exploration,
  author    = {Adamson, Duncan and Spirakis, Paul G.},
  title     = {Brief announcement: Exploration of always {$S$}-connected temporal graphs},
  booktitle = {5th Symposium on Algorithmic Foundations of Dynamic Networks (SAND 2026)},
  series    = {Leibniz International Proceedings in Informatics},
  volume    = {373},
  pages     = {20:1--20:6},
  year      = {2026},
  doi       = {10.4230/LIPIcs.SAND.2026.20}
}

@article{Demaine2015SlidingTokens,
  author  = {Demaine, Erik D. and Demaine, Martin L. and Fox-Epstein, Eli and Hoang, Duc A. and Ito, Takehiro and Ono, Hirotaka and Otachi, Yota and Uehara, Ryuhei and Yamada, Takeshi},
  title   = {Linear-time algorithm for sliding tokens on trees},
  journal = {Theoretical Computer Science},
  volume  = {600},
  pages   = {132--142},
  year    = {2015},
  doi     = {10.1016/j.tcs.2015.07.037}
}

@article{Hoang2024DistanceIndependentSet,
  author  = {Hoang, Duc A.},
  title   = {On the complexity of distance-{$d$} independent set reconfiguration},
  journal = {Theoretical Computer Science},
  volume  = {1009},
  pages   = {114682},
  year    = {2024},
  doi     = {10.1016/j.tcs.2024.114682}
}

@inproceedings{Stern2019MAPF,
  author    = {Stern, Roni and Sturtevant, Nathan R. and Felner, Ariel and Koenig, Sven and Ma, Hang and Walker, Thayne T. and Li, Jiaoyang and Atzmon, Dor and Cohen, Liron and Kumar, T. K. Satish and Boyarski, Eli and Bart{\'a}k, Roman},
  title     = {Multi-agent pathfinding: Definitions, variants, and benchmarks},
  booktitle = {Proceedings of the International Symposium on Combinatorial Search},
  volume    = {10},
  pages     = {151--158},
  year      = {2019},
  doi       = {10.1609/socs.v10i1.18510}
}

@article{Fluschnik2022MultistageVertexCover,
  author  = {Fluschnik, Till and Niedermeier, Rolf and Rohm, Valentin and Zschoche, Philipp},
  title   = {Multistage vertex cover},
  journal = {Theory of Computing Systems},
  volume  = {66},
  pages   = {454--483},
  year    = {2022},
  doi     = {10.1007/s00224-022-10069-w}
}

@article{Chimani2023Multistage,
  author  = {Chimani, Markus and Troost, Niklas and Wiedera, Tilo},
  title   = {A general approximation for multistage subgraph problems},
  journal = {Procedia Computer Science},
  volume  = {223},
  pages   = {334--342},
  year    = {2023},
  doi     = {10.1016/j.procs.2023.08.245}
}

@article{Ibiapina2024SnapshotDisjointness,
  author  = {Ibiapina, Allen and Silva, Ana},
  title   = {Snapshot disjointness in temporal graphs},
  journal = {Theoretical Computer Science},
  volume  = {1009},
  pages   = {114690},
  year    = {2024},
  doi     = {10.1016/j.tcs.2024.114690}
}

@article{Davot2026Reconfiguration,
  author  = {Davot, Tom and Enright, Jessica and Larios-Jones, Laura},
  title   = {Parameterised algorithms for temporally satisfying reconfiguration problems},
  journal = {Discrete Applied Mathematics},
  volume  = {395},
  pages   = {493--508},
  year    = {2026},
  doi     = {10.1016/j.dam.2026.09.006}
}

@inproceedings{Hirahara2025Extended,
  author    = {Hirahara, Shuichi and Ohsaka, Naoto and Suga, Tatsuhiro and Suzuki, Akira and Tamura, Yuma and Zhou, Xiao},
  title     = {Reachability of independent sets and vertex covers under extended reconfiguration rules},
  booktitle = {36th International Symposium on Algorithms and Computation (ISAAC 2025)},
  series    = {Leibniz International Proceedings in Informatics},
  volume    = {359},
  pages     = {39:1--39:20},
  year      = {2025},
  doi       = {10.4230/LIPIcs.ISAAC.2025.39}
}

\end{document}